\documentclass{jaa}
\usepackage{natbib}
\usepackage{newtxtext,newtxmath}
\usepackage{graphicx}
\usepackage{amsmath}
\usepackage{caption}
\usepackage{subcaption}
\usepackage{siunitx}
\usepackage{float}
\usepackage[export]{adjustbox}
\usepackage{graphicx}
\usepackage{aas_macros}

\usepackage[colorlinks=true, allcolors=blue]{hyperref}

\begin{document}\sloppy

\title{$Herschel$ investigation of cores and filamentary structures in L1251 located in the Cepheus flare }


\author{Divyansh Dewan\textsuperscript{1,2,*}, Archana Soam\textsuperscript{2},
Guo-Yin Zhang$^{3}$,
Akhil Lasrado$^{2}$,
Saikhom Pravash Singh$^{2}$,
and Chang Won Lee\textsuperscript{4}}
\affilOne{\textsuperscript{1}Indian Institute of Science Education and Research, Campus Road, Mohanpur, Kolkata 741246, West Bengal, India\\}
\affilTwo{\textsuperscript{2}Indian Institute of Astrophysics, II Block, Koramangala, Bengaluru 560034, India \\}
\affilThree{\textsuperscript{3}National Astronomical Observatories, Chinese Academy of Sciences, Beijing 100101, PR China\\}
\affilFour{\textsuperscript{4}Korea Astronomy and Space Science Institute (KASI)
776 Daedeokdae-ro, Yuseong-gu, Daejeon 34055, Republic of Korea}

\twocolumn[{

\maketitle

\corres{dd20ms116@iiserkol.ac.in}


\begin{abstract}
Context: Molecular clouds are the prime locations of star formation. These clouds contain filamentary structures and cores which are crucial in the formation of young stars. \\
Aims: In this work, we aim to quantify the physical properties of structural characteristics within the molecular cloud L1251 to better understand the initial conditions for star formation. \\
Methods: We applied the \emph{getsf} algorithm to identify cores and filaments within the molecular cloud L1251 using the \emph{Herschel} multiband dust continuum image, enabling us to measure their respective physical properties. Additionally, we utilized an enhanced differential term algorithm to produce high-resolution temperature maps and column density maps with a resolution of ${13.5}''$.\\
Results: We identified 122 cores in the region. Out of them, 23  are protostellar cores, 13 are robust prestellar cores, 32 are candidate prestellar cores (including 13 robust prestellar cores and 19 strictly candidate prestellar cores), and 67 are unbound starless cores. \emph{getsf} also found 147 filament structures in the region. Statistical analysis of the physical properties (mass ($M$), temperature ($T$), size, and core brightness (hereafter, we are using the word luminosity ($L$)) for the core brightness) of obtained cores shows a negative correlation between core mass and temperature and a positive correlation between ($M$/$L$) and ($M$/$T$). Analysis of the filaments gives a median width of 0.14 pc and no correlation between width and length. Out of those 122 cores, 92 are present in filaments (~75.4\%) and the remaining were outside them. Out of the cores present in filaments, 57 (~62\%) cores are present in supercritical filaments ($M_{\rm line}>16M_{\odot }/{\rm pc}$)
\end{abstract}


}]


\doinum{12.3456/s78910-011-012-3}
\artcitid{\#\#\#\#}
\volnum{000}
\year{0000}
\pgrange{1--}
\setcounter{page}{1}
\lp{1}
\section{Introduction}
	Stars are found to be forming in the dense parts of the interstellar medium (ISM) called molecular clouds. Inside these clouds, star formation is found to occur in the large filamentary structures (\citealt{2010A&A...518L.102A, molinari2010clouds}). Using \emph{Herschel} data, it has been found that these dusty structures are hierarchical in nature  (\citealt{men2021multiscale, article}) and contain smaller units such as clumps, cores, filaments, and hub-filament structures. The investigation of the statistics of these structures can provide us with clues on the formation and evolution of star-forming regions.
	
	\emph{Herschel} dust maps at sub-millimeter wavelengths have played a crucial role in showing the ubiquity of filamentary networks present in the cold ISM and their importance in star formation in molecular clouds. Surveys of multiple molecular clouds reveal a striking property of these filaments in that they tend to have a characteristic width of 0.1 pc (\citealt{2011A&A...529L...6A}). The origin of this characteristic width is not yet clear with multiple explanations claiming the width is a result of the thermodynamics of ISM gas (\citealt{10.1111/j.1365-2966.2005.08881.x}), to the magneto-sonic scale below which the turbulence becomes subsonic in diffuse, non-star-forming molecular gases (\citealt{padoan2001turbulent}). It has also been observed that the majority of prestellar cores form in “supercritical” filaments where the mass per unit length exceeds the critical line mass. This is calculated by modeling the filaments as nearly isothermal long cylinders at a temperature of 10 K and is found to be $M_{\rm line}>16M_{\odot }/{\rm pc}$ (\citealt{inutsuka1997production}). These supercritical filaments are prone to gravitational instability, and as a result, they can undergo fragmentation (\citealt{inutsuka1997production}). This fragmentation process leads to the formation of denser regions called cores within the filaments (\citealt{zhang2020fragmentation}).
	
	Recent research into filaments has revealed more details regarding their structure and how it may impact core formation inside them. These include the existence of velocity gradients roughly perpendicular to the axis of the filament. These indicate the accretion of gas from the molecular cloud into the filaments (\citealt{shimajiri2019probing}, \citealt{chen2020velocity}). This accretion flows along the filament in radial direction as shown in \cite{hacar2022initial, zhang2020fragmentation}. The relation of mass growth of cores within these filaments with this accretion was proposed by \cite{gehman1996wave} and has been shown in works like \cite{banerjee2006supersonic} and \cite{gomez2018magnetic}
	
	There is a strong link present between the formation of prestellar cores and filaments to star formation in these dense molecular clouds. These \emph{Herschel} survey results are in remarkable agreement with theory and numerical simulations consistently showing that the ISM is highly filamentary on all scales and star formation is intimately connected with self-gravitating filaments (\citealt{hennebelle2013origin}). These simulations now successfully include turbulence, gravity, various cooling processes, magnetohydrodynamics, and radiation from massive stars. These numerical advances have provided us with new and valuable insights into the
	physics of filaments and star formation, including the formation, fragmentation, and further evolution of filaments through accretion, and the central role of filaments in the
	rapid gathering of gas into cluster-forming, dense regions (\citealt{konyves2015census, li2023properties}).
	
\begin{figure*}
		\centering
		\includegraphics[width=0.7\textwidth]{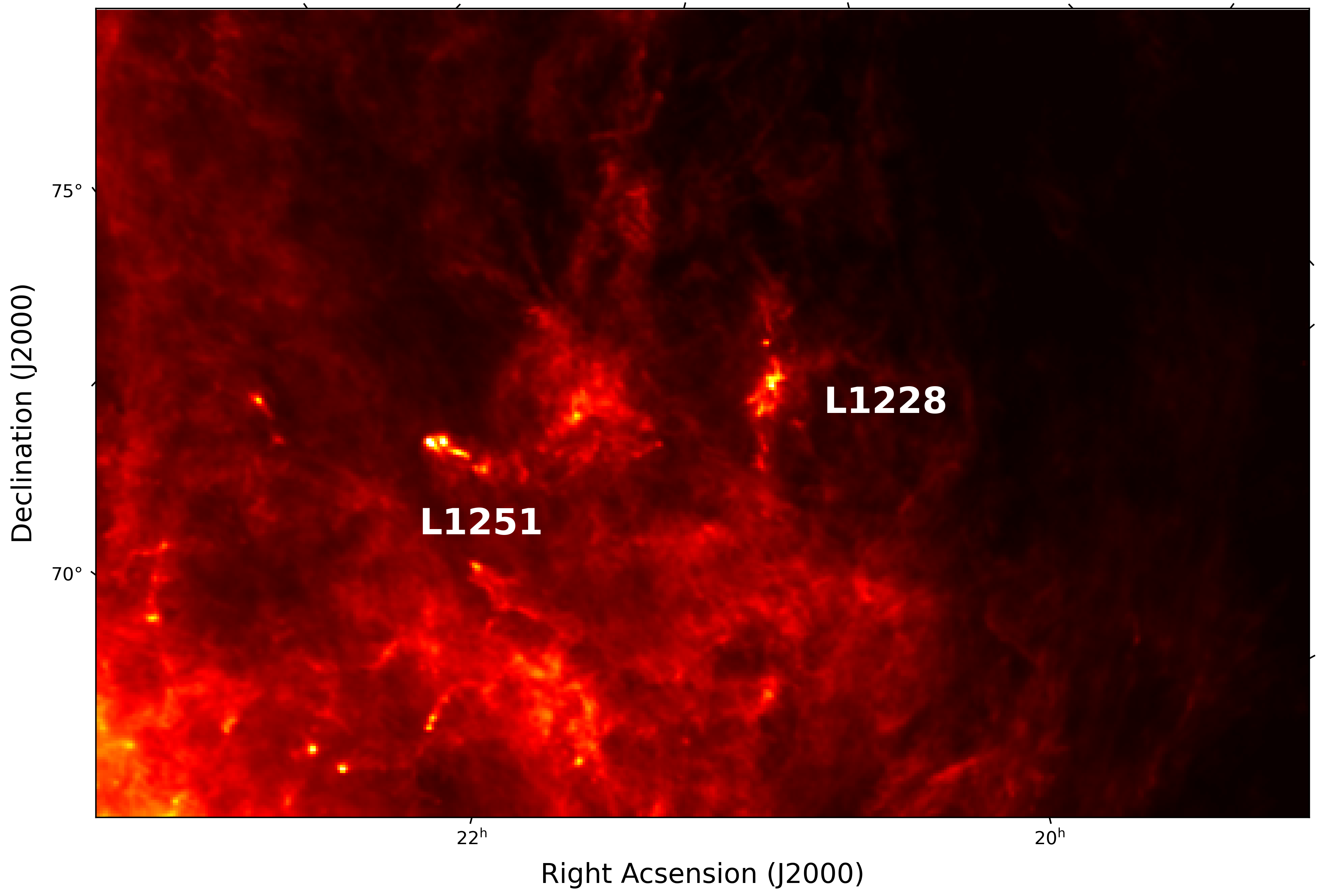}
		\caption{L1251 highlighted in the 857 GHz Plank map of the Cepheus flare along with another molecular cloud L1228}
		\label{flare}  
	\end{figure*}

L1251 is a molecular cloud present in Cepheus Flare at a distance of $\sim$ 340 $\pm$7 pc, extending from $l \sim 100^{\circ} - 116^{\circ}, b \sim 9^{\circ} - 25^{\circ}$ (Right Ascension$\sim 19^{h}-23^{h}$, Declination $\sim 68^{\circ} - 80^{\circ}$)   (\citealt{Sharma_2022}) as shown in Fig \ref{flare} It has an elongated structure with a cometary morphology, which is believed to have been formed as a consequence of its interaction with a supernova bubble (\citealt{1989ApJ...347..231G, Sharma_2022}). The structure of the cloud, shown clearly in Fig \ref{350}, alongside the elongated structures mentioned above is similar to a body flying at hypersonic speed across an ambient medium (\citealt{2004A&A...425..133B}). The position angle of the elongated cloud is found to be $107^{\circ}$.
	
	The objective of this paper is to study the structure of this cloud by identifying the cores and filaments present and studying their physical properties. Our method in extracting cores and filaments is very similar to the work done in \cite{Zhang_2022} where they studied the Perseus molecular cloud. We used multi-wavelength \emph{Herschel} data (\ref{Herschel_data}) to create high-resolution (\SI{13.5}{\arcsecond}) column density and temperature maps using an improved difference-term algorithm. We then extracted cores and filaments using \emph{getsf} along with their physical properties. Understanding the formation of prestellar cores is crucial for unraveling the early stages of star formation. By analyzing the properties of cores and filaments in L1251 using the new tool \emph{getsf}, we may gain insights into the role of these structures in facilitating star formation. This can contribute to broader discussions on the influence of cloud morphology on star formation efficiency. We also get to compare the results of different software used to detect prestellar cores in the region
	This work is similar to existing work, particularly with \cite{2020ApJ...904..172D} and \cite{10.1093/mnras/stw2648}, but while the former study employs other algorithms \emph{getsources} (\citealt{getsources}) and \emph{getfilaments} (\citealt{getfilaments}) for core and filament extraction, respectively, using \emph{Herschel} images of Cepheus Flare in 70, 160, 250, 350, and 500 $\mu$m with pixel sizes of the images set as \SI{3.2}{\arcsecond}, \SI{3.2}{\arcsecond}, \SI{6}{\arcsecond}, \SI{10}{\arcsecond}, and \SI{14}{\arcsecond}
		respectively, we use \emph{getsf} (\citealt{men2021multiscale}), which handles core and filament extraction simultaneously. By comparing our results with theirs, we can see the difference in the results of these algorithms, especially since \emph{getsf} doesn't take any input from the user. Meanwhile, the latter study uses SCUBA-2 observations in 450 and 850 $\mu$m wavelengths with maps on to \SI{6}{\arcsecond} pixels and CSAR algorithm (\citealt{CSAR}); while we use \emph{Herschel} dust maps at 160, 250, 350, and 500 $\mu$m.  We also focus exclusively on L1251 and find correlations between the mass, luminosity, size, and temperature of detected cores.

	In Sect \ref{Section_2}, we discuss the data we have used in our analysis and the algorithm we use for identifying cores and filaments. We then discuss the results including the cores and filaments we extracted, their properties, along with the correlations between them in Sect \ref{Section_3} We also do a small comparison with the findings of \cite{2020ApJ...904..172D}. Finally, in Sect \ref{Section_4}, we conclude our work and list our findings.
	\section{Data, Observation, and Techniques} {\label{Section_2}}
	
	
	\subsection{\emph{Herschel} Data} 
\begin{figure*}
		\centering
		\includegraphics[width=0.7\textwidth]{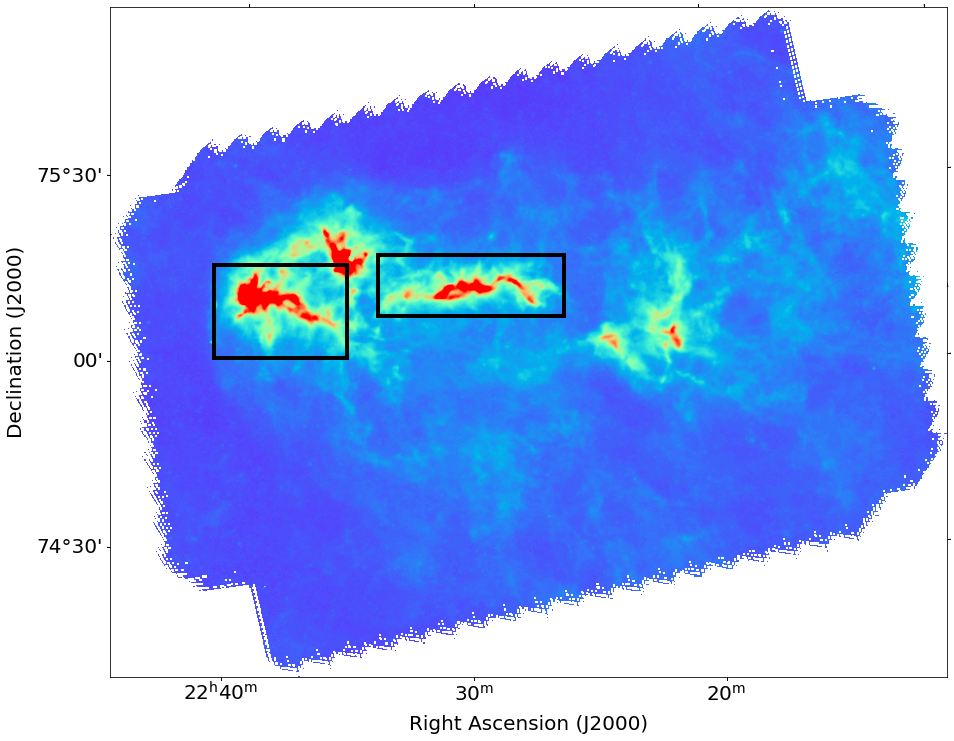}
		\caption{\emph{Herschel} SPIRE 350 $\mu$m image. The elongated structures are highlighted in the picture}
		\label{350}  
	\end{figure*}

{\label{Herschel_data}}
	\label{sec:Herschel data} 
	\emph{Herschel} PACS images in 70 and 160 $\mu$m (\citealt{Poglitsch}) and \emph{Herschel} SPIRE images in 250, 350 and 500 $\mu$m (\citealt{Griffin}) were used in our analysis (Fig \ref{350}). The SPIRE/PACS parallel-mode was used to make scan maps with a speed of \SI{60}{\arcsecond} s$^{-1}$. The beam sizes for \emph{Herschel} observations in 70, 160, 250, 350, and 500 $\mu$m are 8.4, 13.5, 18.2, 24.9, and \SI{36.3}{\arcsecond}, respectively. The data was downloaded from  ESA \emph{Herschel} Science Archive \footnote{http://archives.esac.esa.int/hsa/whsa/}.  The observation id's for the images used are 1342188654, 1342188655, 1342189663, and 1342189664.
	
	We assume that each band of $\it Herschel$ satisfies optically thin dust 
	emission (\citealt{2014A&A...571A..11P})
	\begin{equation}
		I_{\nu}=\kappa_0\left(\nu / \nu_{0}\right)^{\beta} r \mu m_{\rm H} N_{{\rm H_{2}}} B_{\nu}(T) ,
		\label{emissionlaw}
	\end{equation}
	where $I_{\nu}$ is the specific intensity, $\kappa_0$ is the emission 
	cross-section at a reference frequency $\nu_0$, $r$ is the dust-to-gas 
	mass ratio, $\mu$ is the mean molecular weight, $m_{\mathrm{H}}$ is the 
	mass of a hydrogen atom, $N_{\mathrm{H_2}}$ is the gas surface density, 
	$B_{\nu}(T)$ is the Planck function for dust at temperature $T$. 
	
	A power-law form was assumed for the dust opacity\citep{roy2014reconstructing}, 
	$\kappa_{\lambda}=0.1 \times (\lambda  /300\,\mu\rm{m})^{-\beta }\ cm^{2}\,g^{-1}$. \ $\beta$ could vary with frequency, grain size, and grain temperature, but using a fixed value of $\beta=2$ the accuracy of surface density will not deviate by more than 50\% \citep{roy2014reconstructing}. We use Planck data as a reference benchmark and convolve each band of Herschel to the $\it Planck$ resolution. The offset in flux density of each band of $\it Herschel$ can be obtained by linear comparison \citep{molinari2010clouds}. They are 42.9, 11.3, 3.9,  and 0.8 $\rm MJy\ sr^{-1}$ at 160, 250, 350, and 500 $\mu$m, respectively.

\begin{figure*}
		\centering
		\includegraphics[width=0.8\textwidth]{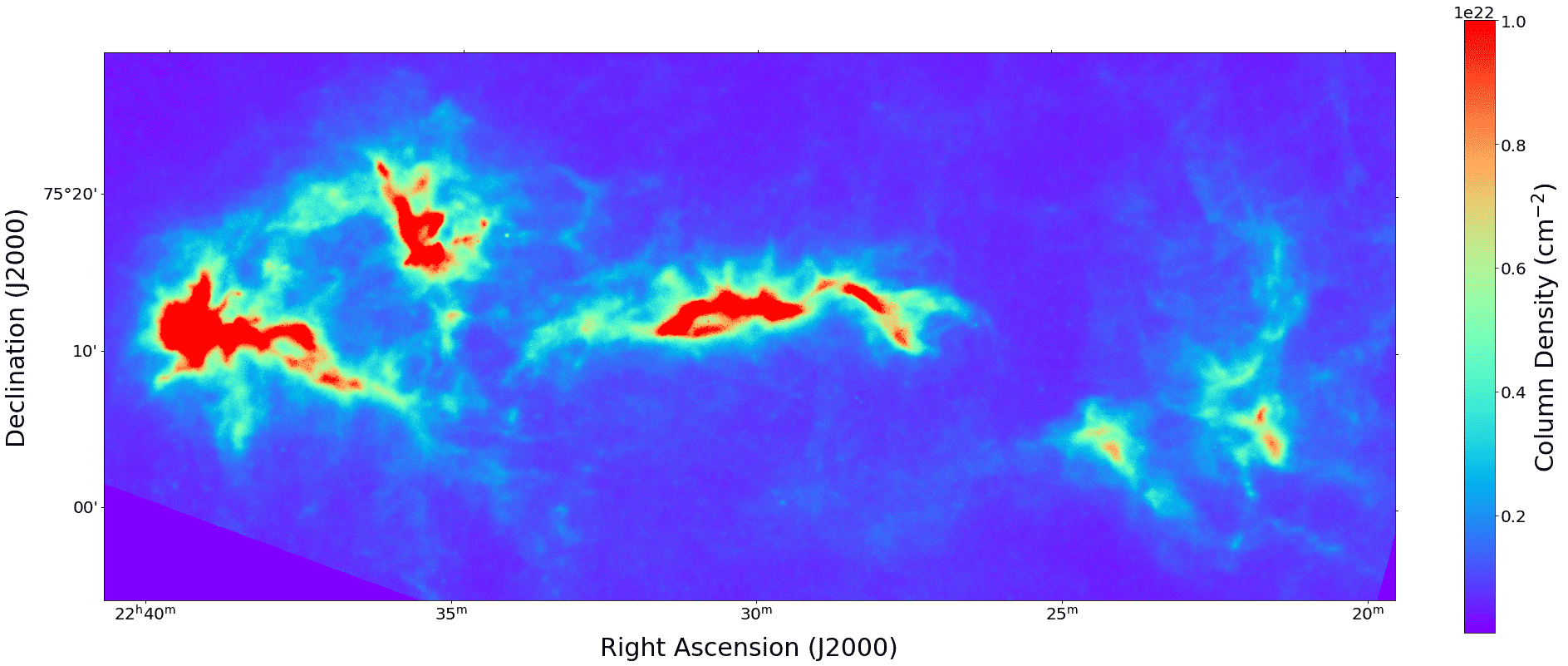}
		\caption{\SI{13.5}{\arcsecond} resolution column density map of L1251 made using \emph{getsf}}
		\label{col}
	\end{figure*}

\begin{figure*}
		\centering
		\includegraphics[width=0.8\textwidth]{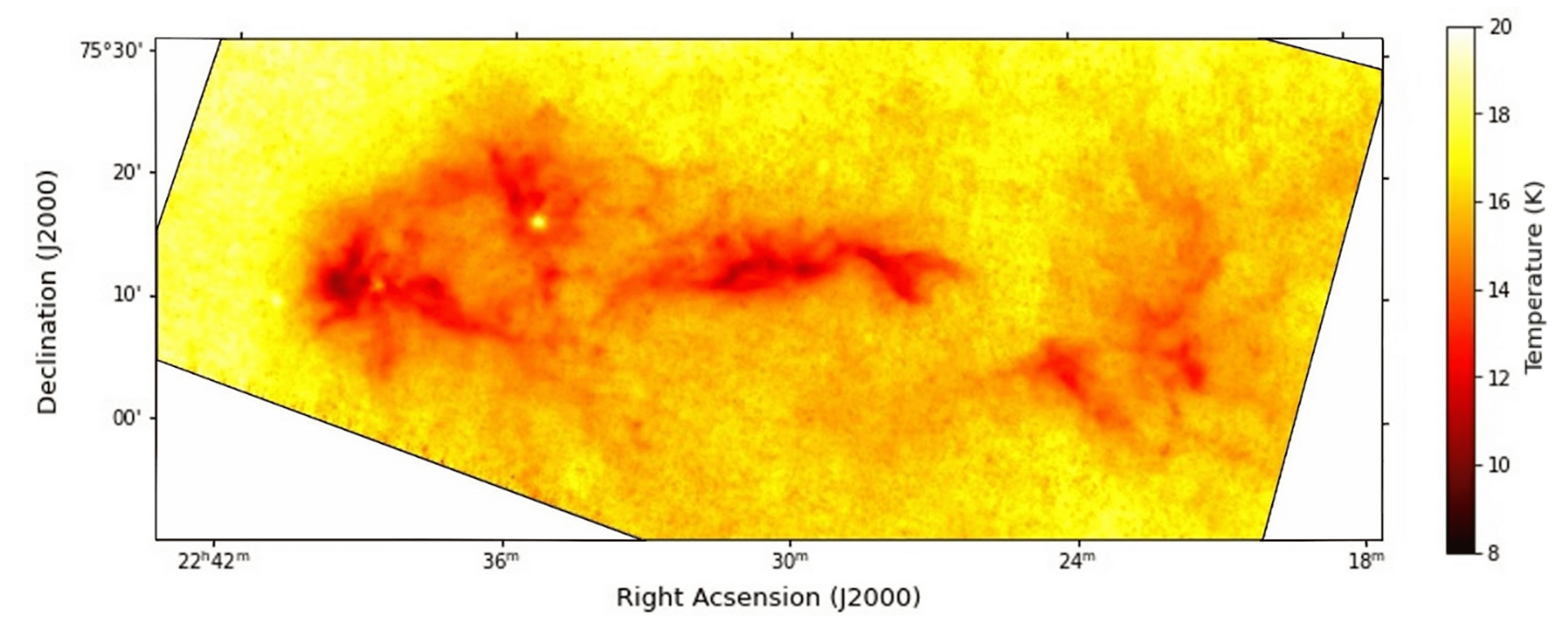}
		\caption{\SI{13.5}{\arcsecond} resolution temperature map of L1251 made using \emph{getsf}}
		\label{temp}
	\end{figure*}

	We also used these multi-wavelength dust maps to obtain high-resolution (\SI{13.5}{\arcsecond}) column density and temperature maps of the region using pixel-by-pixel SED fitting of the data to a modified blackbody function. Instead of convolving all images to the lowest resolution and using that directly for fitting, we used \emph{getsf}, which works using the method described in \cite{men2021multiscale} which employs a difference term algorithm to increase the resolution from \SI{36.3}{\arcsecond} as we would have obtained from the former method to \SI{13.5}{\arcsecond}. The obtained density and temperature maps via \emph{hires} are shown in Fig \ref{col} and \ref{temp} respectively.
	
	We now use the derived column density maps to obtain the total mass ($M$) of L1251. We use the formula  
	\begin{equation}
		M = N(H_{2})\mu m_{H} A, 
		\label{masseqn}
	\end{equation}
	where $N(H_{2})$ is the derived column density value from the column density map Fig \ref{col}, $\mu$ is the mean weight of molecular gas taken to be 2.8 assuming that the gas is 70\% molecular hydrogen by mass (\citealt{2010A&A...518L..92W}), $m_{H}$ is mass of hydrogen atom, and A is projected area. The boundary of the map was chosen by an observational mask where the usable part of the map that contained the cloud was selected. This selection was made empirically. The remaining part was considered background and was excluded from measuring mass. While making the map, \emph{getsf} separates the background image. This way, the mass of the background component is subtracted from the mass of the cloud.
	Using this, we find the mass to be $\approx 2000 \; M_\odot $.
	
	\subsection{\textsc{Getsf}}
	We use \emph{getsf}, a method for extracting sources and filaments in astronomical images using separation of their structural components (\citealt{men2021multiscale}), to find, measure, and extract cores and filamentary structures in L1251. \emph{getsf} identifies structures in astronomical images by separating their structural components. The basic steps involved in using \emph{getsf} and its image processing are described below:
	\begin{enumerate}
		\item First, all the different images need to be resampled to images with the same pixel size and number. This is done via the built-in script - \emph{resample}. An observational mask may also be needed to reduce computational time.
		\item These resampled images are then used to obtain column density and temperature maps using the built-in script - \emph{hires}
		\item Then, \emph{getsf} uses structural decomposition into single-scale images to isolate core and filamentary structures from their background. Noise due to local fluctuations is removed via image flattening.
		\item These cleaned single-scale images are then combined over all the wavelengths given, cores and filaments are identified, and their physical properties are measured and cataloged using scripts \emph{fmeasure} and \emph{smeasure}.
		\item The location and properties of detected structures are obtained in an extraction catalog.
	\end{enumerate}
	
	\emph{getsf} is a nearly fully autonomous algorithm with the only user input needed for \emph{hires} being the offset in flux density (in MJy/sr) for the image, correcting the image absolute calibration; and the maximum size of the structure (in arcsecs) for detecting cores and filaments. More details on the working of \emph{getsf} are in \ref{C}

\section{Results and Analysis} {\label{Section_3}}

\subsection{Core selection and classification}
The obtained cores via \emph{getsf} are selected by the criteria described in \ref{A} This selection is based on benchmark tests given in \cite{men2021multiscale} and is based on properties like their size, detection significance, intensities, etc. 122 cores were identified by this method. Properties like their mass and bolometric luminosity can be found by pixel-by-pixel SED fitting the integrated flux measured at each wavelength by \emph{getsf} to a modified blackbody function. The core size is defined as the mean deconvolved FWHM diameter at the resolution of \SI{13.5}{\arcsecond} of an equivalent elliptical Gaussian source: $R = \sqrt{AFWHM \times BFWHM - 13.5^{2}}$, where AFWHM is the size of the major axis at half maxima and BFWHM is the size of the minor axis at half maxima of the detected core. Using these parameters, we classify the cores into different categories.

The classification is done by comparing the cores to a Bonnor-Ebert sphere (\citealt{1956MNRAS.116..351B, 1955ZA.....37..217E}). The critical mass of a Bonnor Ebert sphere - the maximum mass that can support itself against gravitational attraction is given by,

\begin{equation}
M_{\rm crit}
 = \frac{2.4 R c_{\rm s}^{2}}{G}
\label{mcrit}
\end{equation}

from \citet{1956MNRAS.116..351B, 1955ZA.....37..217E}, where R is the radius of the sphere and $c_{\rm s}$ is the isothermal speed of sound, which is calculated using the SED temperature of each core seperately, and \textit{G} is the gravitational constant.

\begin{figure*}
    \centering
    \includegraphics[width=0.6\paperwidth]{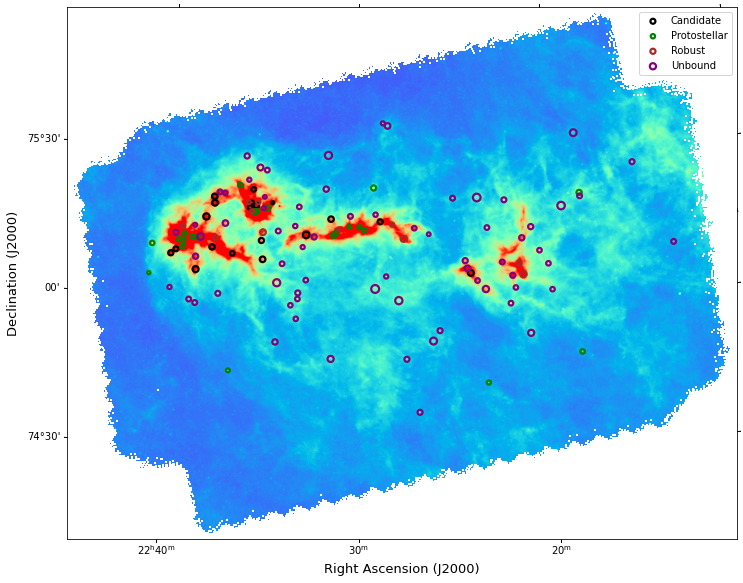}
    \caption{Positions of 122 cores identified via \emph{getsf} overlaid on 250 $\mu$m dust map. 19 strictly candidate prestellar cores are shown in black, 23  protostellar cores are shown in green, 13 robust prestellar cores are shown in red, and 67 unbound starless cores are shown in purple.}
    \label{cores}
\end{figure*}

\begin{enumerate}
    \item The protostars in the region will have infrared emissions. The 70 $\mu$m dust map is better suited to find regions of high temperature caused by these emissions. Hence, our first step is to do a separate \emph{getsf} run with just the 70 $\mu$m image to locate the protostars in the cloud. 
    These cores are then checked via the criteria in \ref{A}. All those who satisfy the criteria are termed protostellar cores. As an additional measure, we have also cross-checked these locations with the SIMBAD database to see if there are any known protostars in the vicinity of identified protostellar cores and also to remove any galaxies erroneously detected. The details of this are given in \ref{B}.

    \item If the ratio $\alpha _{\rm BE}=M_{\rm crit}/M_{\rm core} \leq$ 2, the core is deemed self-gravitating. This is analogous to the criteria used to select self-gravitating objects based on the virial mass ratio (\citealt{bertoldi1992pressure}) Cores meeting this criterion are called robust prestellar cores 
    \item An empirical condition $\alpha_{BE} \leq 5 \times (\sqrt{(AFWHM \times BFWHM)}/\SI{13.5}{\arcsecond})^{0.4}$ obtained by Monte Carlo simulations in \cite{konyves2015census} is used to select candidate prestellar cores. The simulated sources in \cite{konyves2015census} are all gravitationally bound, but the above-mentioned mass criteria cannot fully select them, while this empirical formula can screen out 95\% of the gravitationally bound sources. We have modified the denominator in the formula to 13.5 to account for the resolution of our column density map.
    
    \item The remaining cores are termed as unbound starless.
\end{enumerate}
This methodology of the classification is described in \citet{2015A&A...584A..91K}. Out of 122 total cores, we found that 23 are protostellar, 13 are robust prestellar cores, 32 are candidate prestellar (including 13 robust prestellar cores and 19 strictly candidate prestellar cores), and 67 are unbound starless cores. All of the selected robust prestellar cores also belong to the broader criteria of candidate prestellar cores and hence are also counted with them. The 19 strictly candidate cores are those that do not fit the criteria for robust cores. The obtained cores are marked on the 250 $\mu$m dust map of L1251 in Fig. \ref{cores}.
\\

\subsection{Statistical analysis of core properties}
\begin{figure*}

    \begin{subfigure}{0.4\paperwidth}
        \centering
        \includegraphics[width=0.4\paperwidth]{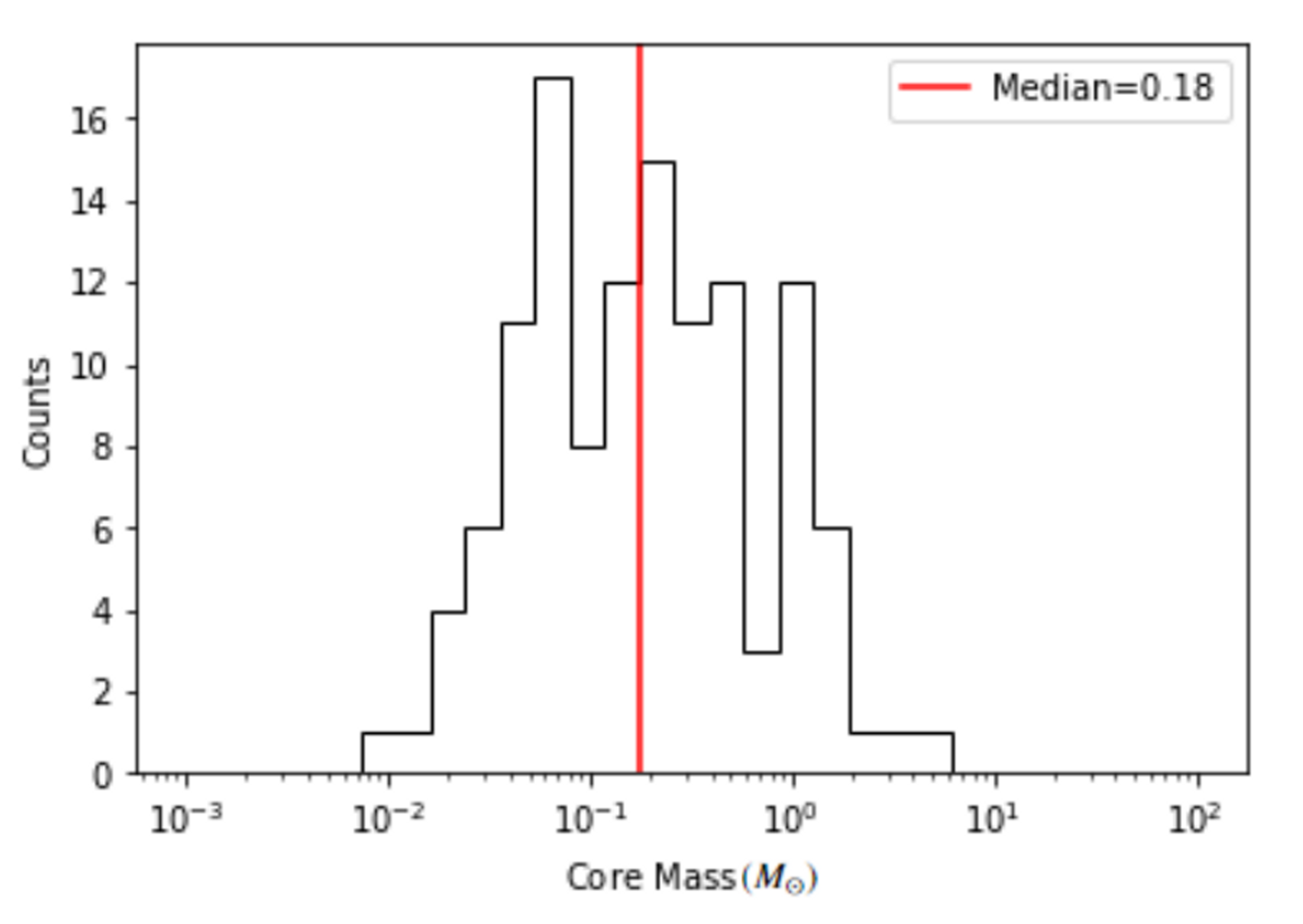}
        \caption{Histogram of masses of detected cores}
        \label{mass_hist}
    \end{subfigure}
    \begin{subfigure}{0.4\paperwidth}
        \centering
        \includegraphics[width=0.4\paperwidth]{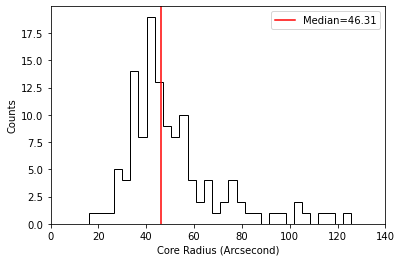}
        \caption{Histogram of radii of detected cores}
        \label{size_hist}
    \end{subfigure}
    \begin{subfigure}{0.4\paperwidth}
        \centering
        \includegraphics[width=0.4\paperwidth]{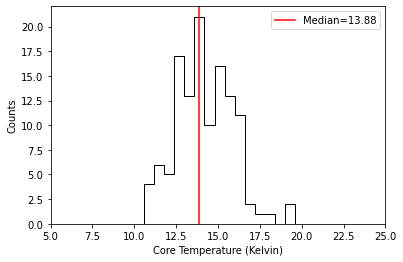}
        \caption{Histogram of temperatures of detected cores}
        \label{temp_hist}
    \end{subfigure}
    \begin{subfigure}{0.4\paperwidth}
        \centering
        \includegraphics[width=0.4\paperwidth]{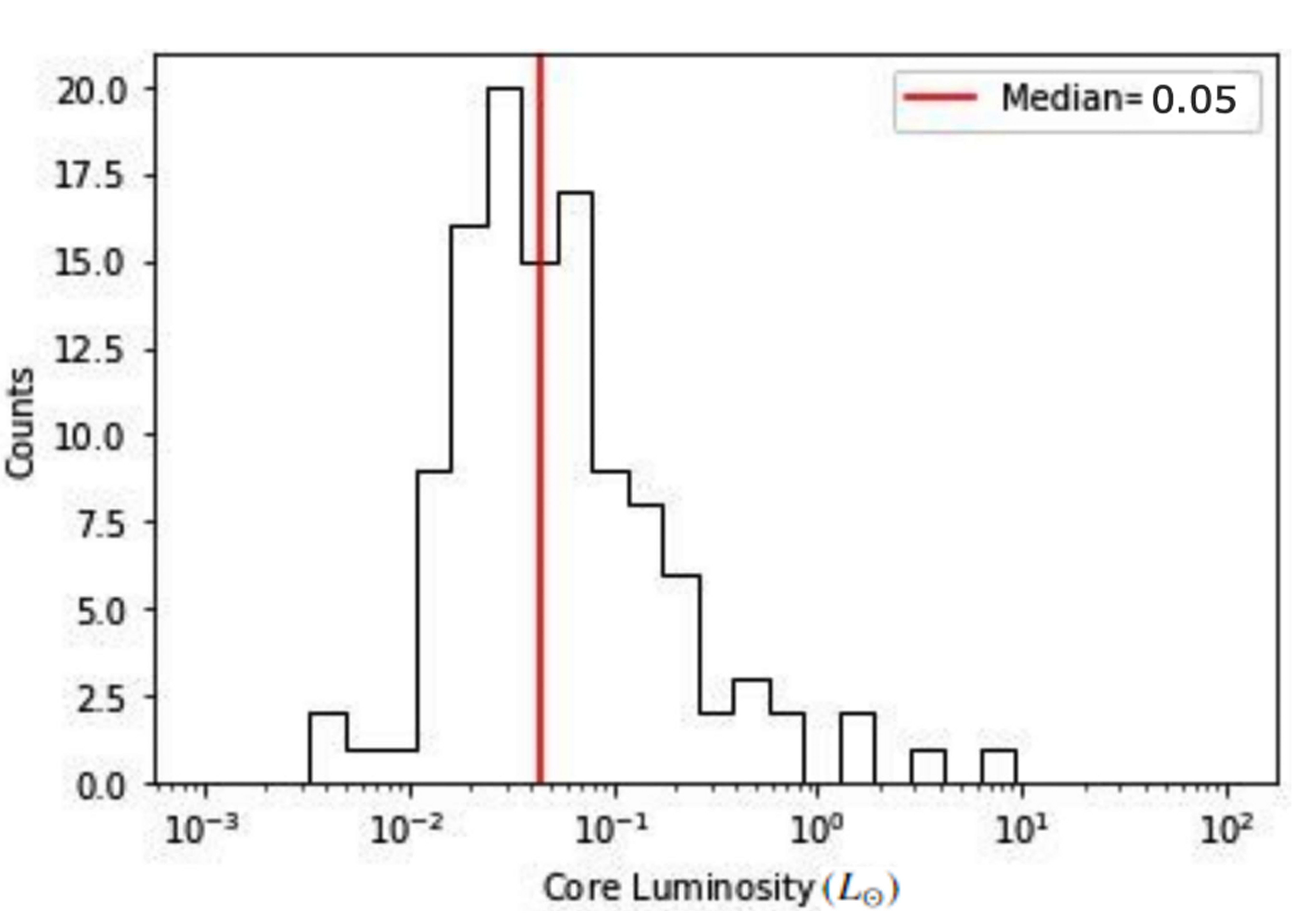}
        \caption{Histogram of luminosities of detected cores}
        \label{lum_hist}
    \end{subfigure}
    \begin{subfigure}{0.4\paperwidth}\vspace{3mm}
    \centering
        \includegraphics[width=0.4\paperwidth]{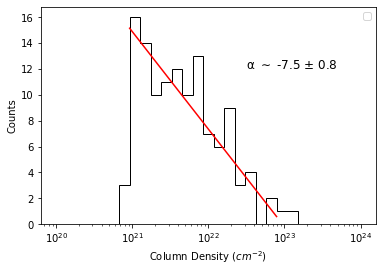}
        \caption{Histogram of column density at locations of cores. The graph is fitted with a power-law of index $\sim$ 7.5 }
         \label{den_hist}
    \end{subfigure}

\caption{Histograms of core properties}
\label{all_hist}
\end{figure*}

Using the data in the \emph{getsf} extraction catalog, the temperature and column density maps, and equation \ref{masseqn} we obtained the mass, size, temperature, and brightness of cores. These properties are shown in the histograms below (Fig \ref{all_hist}). We obtained 0.2 $M_\odot$ as the median mass, \SI{46.3}{\arcsecond} (corresponding to 0.08 pc) as the median radius, 13.9 K as the median temperature, and 0.05 $L_\odot$ as the median luminosity.

\begin{figure*}
    \begin{subfigure}{0.4\paperwidth}
        \centering
        \includegraphics[width=0.4\paperwidth]{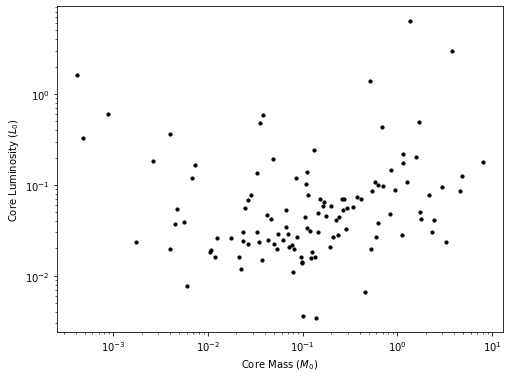}
         \caption{No correlation between mass and luminosity of cores}
        \label{mvsl}
    \end{subfigure}
    \begin{subfigure}{0.4\paperwidth}
        \centering
        \includegraphics[width=0.4\paperwidth]{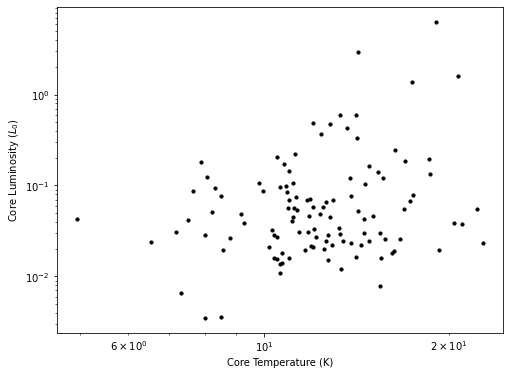}
        \caption{No correlation between luminosity and temperature of cores}
        \label{lvst}
    \end{subfigure}
    \begin{subfigure}{0.4\paperwidth}
        \centering
        \includegraphics[width=0.4\paperwidth]{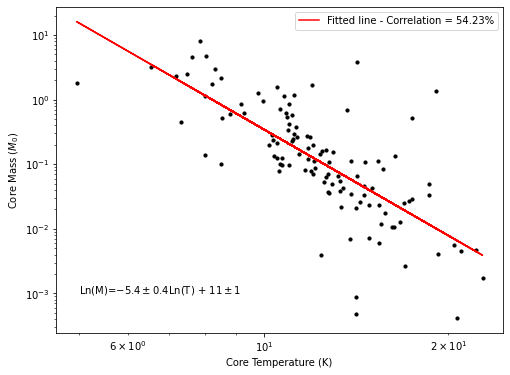}
        \caption{Correlation between mass and temperature of cores}
        \label{corr1}
    \end{subfigure}
    \begin{subfigure}{0.4\paperwidth}
        \centering
        \includegraphics[width=0.4\paperwidth]{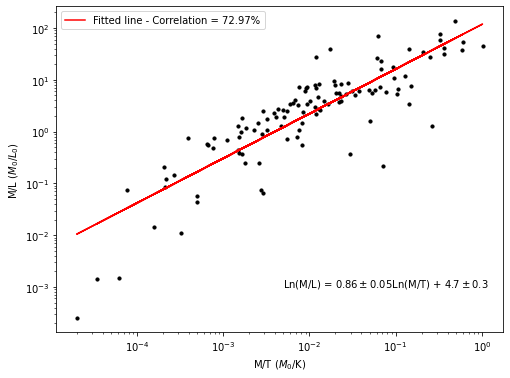}
        \caption{Correlation between ($M$/$L$) and ($M$/$T$) of cores}
        \label{corr2}
    \end{subfigure}
\caption{Correlation between core properties} 
\end{figure*}

We then looked for correlations between these properties. We found no correlation in the way luminosity varied with mass or temperature (Fig \ref{mvsl} and \ref{lvst}) but obtained a negative correlation ($R^{2} = 54.23\%$) of core mass with core temperature (Fig \ref{corr1}) and a positive correlation ($R^{2} = 72.97\%$) between ($M$/$L$) and ($M$/$T$) (Fig \ref{corr2}).

Similiar correlations are also observed by \cite{Zhang_2022} and \cite{10.1093/mnras/stu219}. The negative correlation of core mass with temperature is consistent with our expectations. For a prestellar core, the primary heating source is the Interstellar Radiation Field (ISRF). As the mass of the prestellar core increases, the amount of dust shielding the interior of the core from the radiation increases, and hence, the temperature decreases.

The positive correlation obtained between  ($M$/$L$) and ($M$/$T$) of the cores can be explained by considering the relation of the bolometric luminosity of the cores with their temperature. This relation is shown explicitly in \cite{Zhang_2022}

By making a histogram of the column densities of the positions of cores, we see a sharp jump at a density of $ \sim 10^{21}\mathrm{cm}^{-2} $ after which cores start to form in the cloud. The maximum number of cores are formed just after this density and their number decreases with increasing densities. The histogram can be fitted by using a power-law of index $\sim$ 7.5 as shown in Fig \ref{den_hist}.

\subsection{Detected filaments}
Filaments are complicated in their shapes and widths, often interconnected with each other and various nearby branches, and have variable intensity along their crests. The length of a filament is also not well defined, and neither is it always clear which branch of the complex 3D structure of the filamentary structure belongs to which filament. \emph{getsf} identifies and analyzes filamentary structures by isolating a basic skeleton structure. The simplified skeletons enable an easy selection and better measurements of only the well-behaving, preferably isolated (not blended), and relatively straight parts of the filaments. It identified 147 filament structures in the region. Using the inbuilt script \emph{fmeasure}, we can measure the filamentary structure along its skeleton and find its properties.

\subsection{Statistical analysis of filament properties}
\begin{figure*}
    \centering
    \includegraphics[width=0.6\paperwidth]{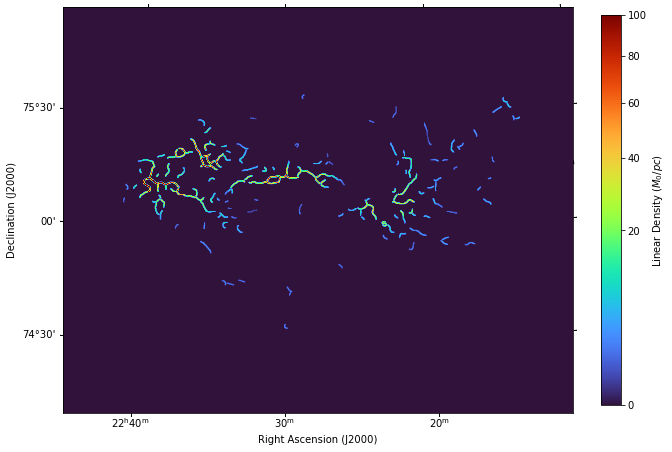}
    \caption{Linear densities of the 147 detected filaments in the region.}
    \label{fil_den}
\end{figure*}

\begin{figure*}
    \begin{subfigure}{0.4\paperwidth}
        \centering
        \includegraphics[width=0.4\paperwidth]{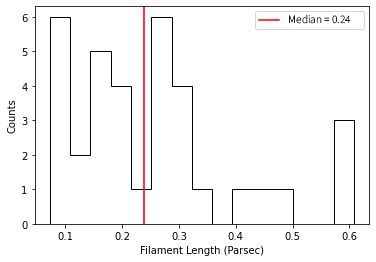}
        \caption{Histogram of lengths of detected filaments}
         \label{fil_len}
    \end{subfigure}
    \begin{subfigure}{0.4\paperwidth}
        \centering
         \includegraphics[width=0.4\paperwidth]{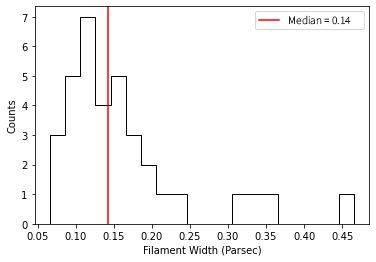}
          \caption{Histogram of widths of detected filaments}
         \label{fil_wid}
    \end{subfigure}
    \begin{subfigure}{0.4\paperwidth}
        \centering
         \includegraphics[width=0.4\paperwidth]{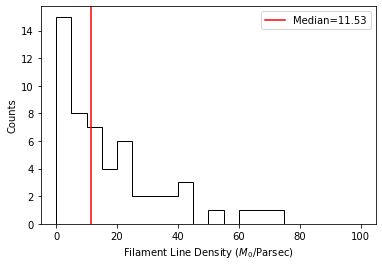}
         \caption{Histogram of densities of detected filaments}
         \label{fil_den2}
    \end{subfigure}
    \begin{subfigure}{0.4\paperwidth}
        \centering
        \includegraphics[width=0.4\paperwidth]{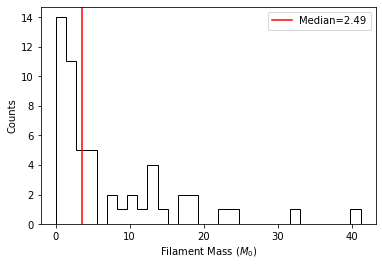}
        \caption{Histogram of masses of detected filaments}
        \label{fil_m}
    \end{subfigure}
\caption{Histograms of filament properties}
\label{fil_hist}
\end{figure*}

Using \emph{fmeasure} to analyze filament properties on the \SI{13.5}{\arcsecond} column density map, we can find physical properties like their lengths, widths, linear densities (same as line mass), and masses. The working of \emph{fmeasure} is described in \ref{C}. Altough the length of a filament is not objectively well defined, it is clear to understand that the longer a segment is, the more reliable it is as a structure to measure filament properties. Out of the 147 identified filaments, we selected the filaments for statistical analysis which had a length greater than 0.2 pc (at least twice the accepted filament width) and detection signal-to-noise ratio $>$2.5. This criteria leaves us with 70 filaments. Out of them, 11 have a length/width ratio $>$ 3. For the selected filaments, we can see the variation in linear densities of the filaments in the region in Fig \ref{fil_den}. From this picture, it is clear that the filaments near the center of the cloud have a higher average density than those near the boundaries with clear regions of high filament densities in the middle and left part of the region. Other properties of the selected filaments are shown in the histograms in Fig \ref{fil_hist}, and \ref{fil_m}. We note that the obtained median width of the identified filaments is $\sim$0.14 pc, in good agreement with the universal characteristic \emph{Herschel} filament width of 0.1 pc  Herschel filaments (\citealt{2011A&A...529L...6A}). The universality of this characteristic width has however been challenged in recent studies, which contend that filament widths are a function of the resolution of the data used to derive them (\citealt{2022A&A...657L..13P}).  

\begin{figure*}
    \centering
    \includegraphics[width=0.6\paperwidth]{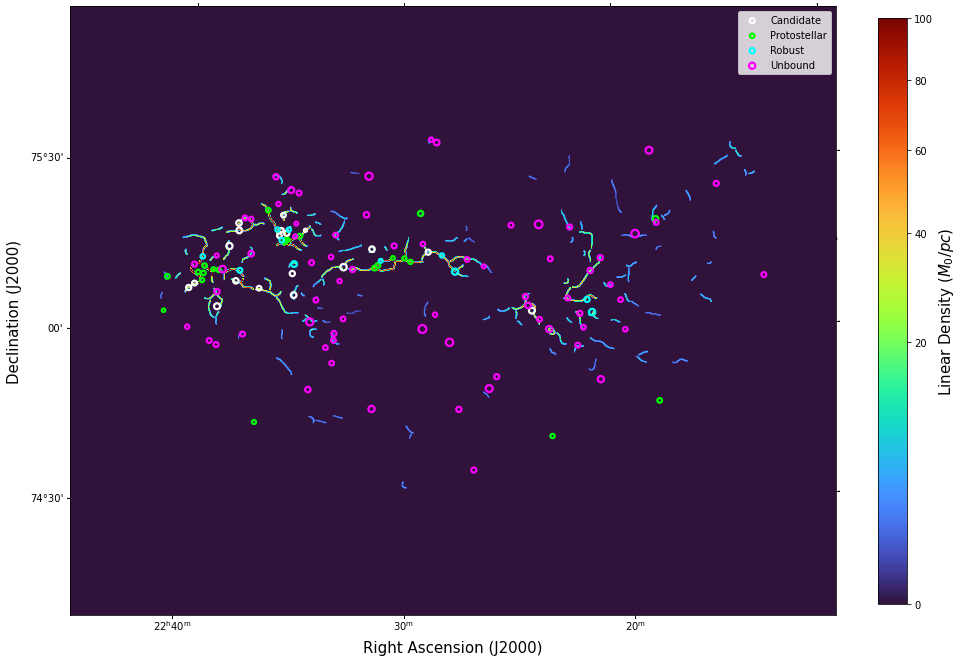}
    \caption{The detected cores overlaid on detected filaments along with their linear densities}
    \label{cores_on_fil}
\end{figure*}

We can now check the position of the cores and see how many of them are located in filaments and the linear density of the filaments they are present in. The methods used to measure linear densities by \emph{getsf} are discussed in \ref{C}. We can calculate the distance 1 pixel corresponds to as the distance to the cloud multiplied by resolution (in radians). Taking the resolution as ${13.5}''$ and distance as 340 pc, we see that 1 pixel corresponds to $\approx 8.9 \times 10^{-9}$ pc. Therefore, 0.05 pc on both sides of the 1-pixel thick filament skeleton (for a total width of 0.1 pc) obtained via \emph{getsf} corresponds to 5.6 pixels on each side. With this, we will claim a core to be present inside a filament if it is within 6 pixels of a filament. We can see from Fig \ref{cores_on_fil} that the majority of cores are present in gravitationally supercritical filaments with linear densities $> 16 M_{\odot}/\mathrm{pc}$. Their distribution with linear density is given in Table \ref{table1}.
\subsection{Comparison with similar work}
If we compare our results with those obtained in \cite{2020ApJ...904..172D} who used \emph{getsources}, an older version of the program \emph{getsf} we used, we see that they obtain a total of 187 cores. Out of them, 11 are protostellar, 53 are robust prestellar, 86 are candidate prestellar and 90 are unbound cores. We can see that using \emph{getsf} on the same region gives us a lower number of cores but a higher number of identified protostellar cores identified via 70 $\mu$m emission. The percentage of unbound cores obtained via \emph{getsf} is 54.92 \% while its 48.13\% using \emph{getsources}
\\
They also found an extensive filamentary network in L1251 and observed 80-100\% of cores to be in filaments, in agreement with our findings
\begin{table*}
\tabularfont
\centering
\begin{tabular}{ccccc} 
	\hline
	\textbf{Linear Density (D) ($M_{\odot}$/pc)} & \textbf{Unbound Starless} & \textbf{Candidate Prestellar} & \textbf{Robust Prestellar} & \textbf{Protostellar} \\
	\hline
	\textbf{Not in Filament} & 25 (37.31 \%) & 0 (0 \%) & 0 (0 \%) & 5 (21.74 \%) \\
	\hline
	\textbf{D $<$ 8} & 18 (26.87 \%) & 0 (0 \%) & 0 (0 \%) & 1 (4.35 \%) \\
	\hline
	\textbf{8 $<$ D $<$ 16} & 8 (11.94 \%) & 7 (21.88 \%) & 2 (15.38 \%) & 1 (4.35 \%) \\
	\hline
	\textbf{D $>$ 16} & 16 (23.88 \%) & 25 (78.13 \%) & 11 (84.62 \%) & 16 (69.56 \%) \\
	\hline
\end{tabular}
\caption{Number of cores in filaments of varying linear densities}
\label{table1}
\end{table*}


\vspace{2em}
\section{Conclusion} {\label{Section_4}}
We studied the dust properties of molecular cloud L1251 in the Cepheus flare region using multiband \emph{Herschel} maps. We began with understanding the cloud structure and other physical properties by making temperature and column densities maps via the \emph{hires} script of \emph{getsf}. \emph{getsf} was also utilised to locate and classify prestellar cores and filaments in the cloud. Using this, we identified 122 cores and 147 filaments. We then used \emph{getsf}'s in-built scripts \emph{smeasure}, \emph{fitfluxes}, and \emph{fmeasure} to inspect the physical properties of cores, namely temperature, size, mass, and luminosity, as well as filament masses, lengths, linear densities, and widths.  

Analysis of this data revealed that out of 122 cores, 92 are present in filaments (~75.4\%) and the remaining do not have an identified parent filament. Of the cores present in filaments, 57 (~62\%) are present in supercritical filaments ($M_{\rm line} > 16 M_\odot/\mathrm{pc}$). We also find a clear correlation between $M$ vs $T$ and ($M$/$L$) vs ($M$/$T$) of the detected cores. 

The former correlation is explained by considering the shielding effect of dust grains on the core from interstellar radiation while the latter correlation is a result of the relation of bolometric luminosity on temperature via Wein's Law. 

Our results regarding core and filament properties are consistent with other studies for \emph{Herschel} filaments of star-forming regions showing a characteristic filament width of $\sim$ 0.1 pc and most prestellar cores existing within dense filaments. We also see a stark column density threshold above which prestellar cores are found (\citealt{Andr__2017}). 
\appendix
\section{Brief description of the working of \emph{getsf}} \label{C}
We used the inbuilt script \emph{hires} of \emph{getsf} to obtain our column density and temperature maps. To make column density and temperature maps, \emph{getsf} uses pixel-by-pixel SED fitting to the Herschel data with a modified blackbody function. 
It produces a base map of the region by convolving all the provided \emph{Herschel} maps at 160, 250, 350, and 500 $\mu$m to the lowest resolution (In our case, \SI{36.3}{\arcsecond}) and using it to create column density and temperature maps via fitting to the blackbody function. Then the 160-350 $\mu$m maps are convolved and fitted to create a less accurate map at \SI{24.9}{\arcsecond} resolution. This map is then convolved to the lower dimension (\SI{36.3}{\arcsecond}) and the difference between this map and the base map is found. This process is repeated for all images at all resolutions. The difference terms are then added to the base density map.
 \\
To identify and select cores and filaments, \emph{getsf} performs the following steps :
\begin{enumerate}
    \item Multi-wavelength images resampled to the same pixel size and the same grid of pixels are taken as input.
    \item The images are spatially decomposed into single-scale images. Separation of the structural components of sources and filaments from each other and from their backgrounds happens from these spatially decomposed images.
    \item The residual noise and background fluctuations in the images of the separated components of sources and filaments are removed via image flattening
    \item These cleaned single-scale images are then combined over all the wavelengths
    \item sources (positions) and filaments (skeletons) are detected in the combined images of the components in their spatially decomposed images
\end{enumerate}  
The properties of cores and filaments are measured and multi-wavelength output catalogs are formed using inbuilt scripts \emph{smeasure} and \emph{fmeasure}.
\emph{getsf} identifies several skeletal structures of the filamentary network and considers the remaining network as its branches. \emph{fmeasure} takes in input as the background-subtracted column density map and uses 2 methods to get the linear densities of filaments, one is by directly integrating the area between the skeleton and the largest extent on each side, and then dividing by length to get density. The other method takes various sampling points along a crest and considers its density. The median value of all sampling points is taken as the density of this filamentary structure. The method is described in full in \cite{men2021multiscale}
\section{Selection criteria for source selection after \emph{getsf} extraction} 
\label{A}
The catalog obtained after a \emph{getsf} extraction contains information on the detected courses. This information is used to select reliable cores from all the detections. The information includes their detection significance and goodness which are parameters relating to signal-to-noise ratio and source reliability, the background subtracted peak intensities of the detected cores, and the major size of half-maximum and the minor size of half-maximum of the detections.

The selection criteria used to select reliable cores is based on benchmark tests from \cite{men2021multiscale}:

\begin{itemize}
    \item $|$GOODM$|$  $>$ 1, where GOODM is monochromatic goodness \\
    \item $|$SIGNM$|$ $>$ 1, where SIGNM is the detection significance from monochromatic single scales. \\
    \item $FXP_{BST}$/$FXP_{ERR}$ $>$ 2, where $FXP_{BST}$ is peak intensity and $FXP_{ERR}$ is peak intensity error. \\
    \item $FXT_{BST}$/$FXT_{ERR}$ $>$ 2, where $FXT_{BST}$ is total flux and $FXT_{ERR}$ is total flux error \\
    \item AFWHM/BFWHM $<$ 2, where AFWHM and BFWHM are the major and minor axis full widths at half-maximum of the elliptical approximations. \\
    \item FOOA/AFWHM $>$ 1.15, where FOOA is full major axis of an elliptical footprint.
\end{itemize}
The first 4 conditions ensure that detected sources are well distinguished from their background. The other 2 are conditions on their size and shape with the 5th condition ensuring the core is circular or elliptical in shape, and the last condition removing cores with unrealistically small ratios of their footprint and half-maximum sizes
\vspace{-2em}
\section{Matches found by cross-referencing protostellar cores with SIMBAD} \label{B}
We cross-referenced the positions of the protostellar cores detected via \emph{getsf} with the SIMBAD database or young stellar objects or cores. Out of 23 protostellar detections, 14 were found to have corresponding SIMBAD entries lying within \SI{10}{\arcsecond}. Their positions and SIMBAD identifiers are tabulated in Table \ref{table2}:
\begin{table*}
\tabularfont
\begin{center}
\begin{tabular}{cccc} 
\hline
\textbf{No.}. & \textbf{RA (J2000)} & \textbf{DEC (J2000)} & \textbf{SIMBAD Identifier}  \\ \hline
1	& 338.8502558	& 75.2851396	&IRAS 22343+7501 \\ 
2	& 337.4295695	& 75.2242914	& [KBB2008] L1251A C2 \\  
3	& 337.8064044	& 75.2133953	& [PWK2017] 62 \\ 
4	& 337.844001	& 75.2057197	& [PWK2017] 63 \\ 
5	& 339.8243047	& 75.1591029	& [PWK2017] 65 \\ 
6	& 339.8120191	& 5.1800094	    & [PWK2017] 72  \\ 
7	& 339.0847361	& 75.3707067	& [PWK2017] 96 \\ 
8	& 338.884616	& 75.2782663	& 2MASS J22352664+7516369 \\  
9	& 339.80096	    & 75.2017797	& 2MASS J22391329+7512161 \\ 
10	& 337.4995097	& 75.2344617	& IRAS 22290+7458 \\ 
11	& 340.2499785	& 75.0652508	& IRAS 22398+7448 \\ 
12	& 339.8728992	& 75.1822671	& JCMTSE J223931.5+751055 \\  
13	& 337.6335849	& 75.2360765	& NAME LDN 1251 A IRS 3 \\ 
14	& 339.6975257	& 75.1921614	& SSTgbs J2238469+7511337 \\ 
\hline
\end{tabular}
\caption{SIMBAD objects corresponding to identified protostars}
\label{table2}
\end{center}
\end{table*}

\section*{Acknowledgements}
We would like to thank Alexander Men'shchikov for his repeated assistance in downloading and using \emph{getsf} which made most of this work possible. The Herschel data was obtained from the ESA's Herschel Science Archive (HSA). DD and AS thank IIA VSP program for providing internship opportunity and supporting this work.
\vspace{-1em}

\bibliography{main} 

\end{document}